# Measuring anomalies in cigarette sales by using official data from Spanish provinces: Are there only the anomalies detected by the Empty Pack Surveys (EPS) used by Transnational Tobacco Companies (TTCs)?

Pedro Cadahia[1], Antonio A. Golpe*[2], Juan M. Martín Álvarez[3].

**Abstract**

Background: There is literature that questions the veracity of the studies commissioned by the transnational tobacco companies (TTC) to measure the illicit tobacco trade. Furthermore, there are studies that indicate that the Empty Pack Surveys (EPS) ordered by the TTCs are oversized. The novelty of this study is that, in addition to detecting the anomalies analyzed in the EPSs, there are provinces in which cigarette sales are higher than reasonable values, something that the TTCs ignore.

Objective: This study analyzed simultaneously, firstly, if the EPSs established in each of the 47 Spanish provinces were fulfilled. Second, anomalies observed in provinces where sales exceed expected values are measured.

Methods: To achieve the objective of the paper, provincial data on cigarette sales, price and GDP per capita are used. These data are modeled with machine learning techniques widely used to detect anomalies in other areas.

Results: The results reveal that the provinces in which sales below reasonable values are observed (as detected by the EPSs) present a clear geographical pattern. Furthermore, the values provided by the EPSs in Spain, as indicated in the previous literature, are slightly oversized. Finally, there are regions bordering other countries or with a high tourist influence in which the observed sales are higher than the expected values.

Conclusions: These results are important because they show that cigarette sales in Spain are conditioned by the effect of tourism and by the price differential with border countries. Along these lines, cooperation between countries in tobacco control policies can have better effects than policies developed based on information from a single country. The lack of control over the transactions of tourists and inhabitants of border countries can cause important anomalies that distort the vision that governments have on tobacco consumption based on official data.

1. Department of Economics. University of Huelva (Spain).
2. Department of Economics. University of Huelva (Spain).
3. Department of Quantitative Analysis. International University of La Rioja (Spain).

*Correspondence to:* Antonio A. Golpe, Department of Economics, University of Huelva, Plaza de la Merced, 11, 21002, Huelva, Spain. *Te*l: +34-959217832. *E-mail address:* antonio.golpe@dehie.uhu.es.

# 1. Introduction

Some theoretical and empirical works have questioned the Empty Pack Surveys (EPSs) because they are commissioned by the transnational tobacco companies (TTCs) and their methodology and validity are not certain[1]. In this context of non-independence of the EPS, it has generated a multitude of papers that have analysed the relationship between what the TTCs show regarding the illicit tobacco trade (ITT) and the official data published by the governments. In addition to the EPSs, the TTCs make reports, usually annually, about ITT. In this line, some studies conclude that the reports made by TTCs require greater transparency, external scrutiny, and the use of independent data[2]. Another issue criticized by some studies is the funding and dissemination by ITT's research TTCs through corporate social responsibility initiatives. In this context, a study concludes that if TTCs data on ITT cannot meet the standards of accuracy and transparency established by high-quality research publications, a solution may be to tax the TTCs and administer the resulting funds to experts independent of the tobacco industry, using previously developed reliable models to measure ITT[3, 4].

In this context of non-independence, many studies have proposed methodologies, using official data, to measure the illicit tobacco market[5]. In this part of the literature there are many results achieved. Some studies conclude that industry-funded estimates inflate likely levels of illicit cigarette use[6-7]. Other papers indicate that industry warnings against tax increases, based on illicit trade rates, in certain countries are not justified[8-10].

Given this academic trend that doubts about the suitability of using EPSs as an indicator of illicit tobacco trade, there are many studies that have focused on analyzing how illicit trade impacts the health of the population, as well as the policies implemented by the governments. Some studies suggest that tobacco tax policies to control the prevalence of smoking and national health care expenditures should be accompanied by a greater effort to curb smuggling activities across borders[11]. On the other hand, some studies have analyzed the impact on the illicit trade of plain tobacco packaging[12-13]. Finally, there are some studies that have carried out a kind of EPSs parallel to the one commissioned by the TTCs to verify its veracity[14-15].

Although there are many studies that have made an effort to contrast the EPSs with official data or by conducting parallel surveys, all of them have focused on analyzing whether the data provided by the TCCs regarding the rates of illicit trade are true. However, a recent study indicates that actual smoking prevalence sometimes exceeds the estimated actual consumption derived from aggregated data on official sales[16]. This means that there are border areas in which the prevalence of smoking is underestimated because official data do not consider what smokers buy in areas with an attractive price differential. In this way, other work also indicates that the excessive production of cigarettes suggests a possible excess supply of cigarettes in some countries, probably diverted towards illicit trade[17]. Studies that contrast the veracity of EPSs focus on verifying whether the rates of illicit trade in a given country are real. However, what is indicated in the works cited in this paragraph highlights the need to also study excess sales rates of products at the borders of certain countries, which can be sold illegally in other countries with an attractive price differential.

Thus, to simultaneously study the veracity of EPSs and excess sales in border areas with another country that has a higher tobacco price, it is necessary to study a country with border countries with lower and higher prices. Spain is a border country with France and Gibraltar, two countries with which it maintains a price differential by excess and by default. In addition, a recent study indicate that distortions are observed in the borders of Spain with France and Gibraltar[18-19]. Although the cited study indicates that there are distortions in the border provinces with France and Gibraltar, it is important to know the magnitude of these distortions. The current health crisis caused by COVID-19 caused that the borders in Spain were closed during the month of April 2020, the effect on cigarette sales being shown in Figure 1. As can be seen, in some provinces sales decreased by up to 180 percent, while in other provinces tobacco sales did not decline. Therefore, focusing the study on Spain seems reasonable if we want to simultaneously analyze the veracity of EPSs and excess sales in border areas.

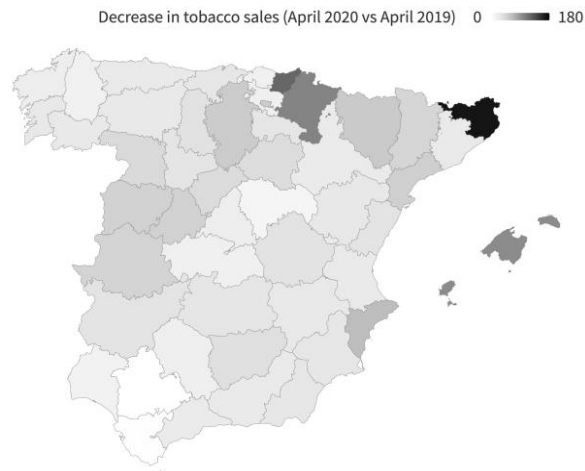

Figura 1. Year-on-year drop in tobacco sales (April 2019 - April 2020).

In this context, our study analyzed the two components of the anomalies in official tobacco sales, that is, both the provinces in which official sales are lower than expected and those in which sales are higher than fair value. There is no evidence in the literature that clarifies the territorial anomalies that are observed in both directions. To the best of our knowledge, this study is the first to analyse, simultaneously, whether the provisions of the EPSs are fulfilled, contrasting it with official data and, furthermore, which provinces show sales above reasonable values.

## 2. Data and Methodology

### 2.1 Data

Our empirical analysis was developed using a panel of data from the Spanish provinces from 2002 to 2017 - last data of the provincial GDP published correspond to the year 2017 -. For cigarette consumption, we used the annual tobacco official sales and the average price of a pack of 20 cigarettes in euros, as published by the Commission for Trade of the Tobacco. The real Gross Domestic Product (GDP) is available in the National Institute of Statistics from Spain. All series employed here are per capita (18 years or older), expressed in real terms using the consumer price index (CPI base 2016).

**Table 1.** Descriptive statistics of the data used.

| Province | Years | Per capita cigarette sales* | | | | | Price* | | | | | Per capita GDP* | | | | |
|---|---|---|---|---|---|---|---|---|---|---|---|---|---|---|---|---|
| | | Mean | SD | Quartile | | | Mean | SD | Quartile | | | Mean | SD | Quartile | | |
| | | | | Q1 | Q2 | Q3 | | | Q1 | Q2 | Q3 | | | Q1 | Q2 | Q3 |
| Albacete | 16 | 93,45 | 25,84 | 64,12 | 101,18 | 115,86 | 3,07 | 1,19 | 1,93 | 2,93 | 4,39 | 19,72 | 3,34 | 18,01 | 21,27 | 21,47 |
| Alicante | 16 | 135,80 | 55,05 | 78,74 | 127,64 | 184,19 | 3,10 | 1,15 | 2,02 | 2,93 | 4,36 | 19,75 | 2,21 | 19,37 | 20,19 | 21,01 |
| Almería | 16 | 114,20 | 38,21 | 72,19 | 118,10 | 146,96 | 3,14 | 1,17 | 2,03 | 2,98 | 4,41 | 21,59 | 2,59 | 20,89 | 21,92 | 23,40 |
| Álava | 16 | 81,18 | 22,34 | 56,98 | 84,07 | 101,80 | 3,06 | 1,21 | 1,92 | 2,91 | 4,37 | 36,26 | 6,96 | 32,92 | 38,29 | 40,87 |
| Asturias | 16 | 88,74 | 21,88 | 64,93 | 95,44 | 107,41 | 3,06 | 1,19 | 1,91 | 2,90 | 4,31 | 20,90 | 3,56 | 19,37 | 22,41 | 23,12 |
| Ávila | 16 | 93,47 | 25,53 | 65,35 | 99,97 | 115,41 | 3,08 | 1,20 | 1,94 | 2,92 | 4,37 | 18,93 | 3,27 | 16,96 | 20,30 | 21,09 |
| Badajoz | 16 | 98,68 | 28,85 | 66,72 | 110,11 | 123,38 | 3,04 | 1,20 | 1,87 | 2,91 | 4,29 | 17,53 | 2,96 | 16,05 | 18,74 | 19,34 |
| Islas Baleares | 16 | 168,33 | 72,05 | 95,93 | 150,77 | 228,62 | 3,13 | 1,15 | 2,05 | 2,93 | 4,37 | 27,29 | 3,66 | 26,06 | 28,11 | 29,15 |
| Barcelona | 16 | 87,59 | 27,11 | 58,16 | 89,84 | 109,03 | 3,04 | 1,20 | 1,89 | 2,86 | 4,34 | 28,75 | 5,25 | 26,19 | 30,21 | 31,28 |
| Vizcaya | 16 | 77,93 | 18,21 | 58,77 | 81,43 | 92,03 | 3,06 | 1,21 | 1,92 | 2,91 | 4,37 | 28,92 | 5,62 | 25,51 | 30,90 | 31,83 |
| Burgos | 16 | 87,44 | 23,79 | 61,89 | 91,81 | 109,79 | 3,05 | 1,21 | 1,89 | 2,88 | 4,33 | 26,61 | 4,62 | 24,13 | 28,33 | 29,22 |
| Cáceres | 16 | 99,28 | 26,80 | 69,04 | 109,19 | 121,22 | 3,04 | 1,19 | 1,89 | 2,90 | 4,34 | 17,25 | 3,14 | 15,54 | 18,33 | 18,86 |
| Cádiz | 16 | 75,60 | 33,69 | 37,81 | 82,92 | 107,68 | 3,04 | 1,19 | 1,89 | 2,90 | 4,32 | 18,82 | 2,40 | 18,64 | 19,65 | 20,27 |
| Cantabria | 16 | 93,99 | 30,05 | 65,44 | 102,02 | 118,87 | 3,08 | 1,20 | 1,93 | 2,91 | 4,41 | 22,51 | 3,54 | 20,86 | 23,99 | 24,55 |
| Castellón | 16 | 102,53 | 33,67 | 66,20 | 103,11 | 133,72 | 3,07 | 1,17 | 1,94 | 2,92 | 4,36 | 25,61 | 3,68 | 24,76 | 26,19 | 27,02 |
| Ciudad Real | 16 | 96,62 | 26,99 | 66,04 | 105,98 | 120,09 | 3,06 | 1,20 | 1,91 | 2,91 | 4,37 | 20,66 | 3,31 | 19,17 | 21,80 | 22,65 |
| Córdoba | 16 | 88,33 | 32,20 | 50,46 | 98,48 | 115,96 | 3,03 | 1,21 | 1,88 | 2,88 | 4,35 | 17,91 | 2,97 | 16,83 | 18,99 | 19,45 |
| La Coruña | 16 | 81,50 | 20,58 | 58,94 | 87,07 | 98,53 | 3,05 | 1,20 | 1,89 | 2,89 | 4,35 | 21,68 | 4,30 | 19,17 | 23,58 | 23,98 |
| Cuenca | 16 | 98,42 | 25,96 | 68,50 | 106,47 | 121,25 | 3,08 | 1,20 | 1,94 | 2,93 | 4,42 | 20,53 | 3,97 | 18,54 | 21,77 | 22,65 |
| Guipúzcoa | 16 | 146,87 | 51,79 | 95,44 | 144,96 | 193,39 | 3,03 | 1,21 | 1,85 | 2,86 | 4,32 | 31,07 | 5,47 | 28,36 | 33,18 | 33,92 |
| Gerona | 16 | 267,40 | 97,35 | 169,15 | 261,90 | 358,11 | 3,05 | 1,20 | 1,87 | 2,86 | 4,36 | 29,22 | 4,39 | 27,97 | 30,58 | 31,27 |
| Granada | 16 | 99,06 | 31,07 | 63,37 | 105,28 | 126,55 | 3,08 | 1,19 | 1,94 | 2,93 | 4,38 | 18,28 | 2,97 | 17,18 | 19,32 | 20,06 |
| Guadalajara | 16 | 93,49 | 28,64 | 61,79 | 96,66 | 116,66 | 3,08 | 1,18 | 1,97 | 2,92 | 4,38 | 20,93 | 2,69 | 20,53 | 21,98 | 22,47 |
| Huelva | 16 | 113,75 | 41,13 | 65,68 | 125,89 | 150,66 | 3,04 | 1,19 | 1,88 | 2,89 | 4,31 | 19,40 | 2,70 | 18,84 | 19,91 | 21,07 |
| Huesca | 16 | 116,51 | 33,40 | 79,16 | 124,19 | 147,17 | 3,07 | 1,20 | 1,92 | 2,90 | 4,35 | 26,90 | 5,23 | 23,35 | 28,87 | 30,21 |
| Jaén | 16 | 95,73 | 27,76 | 64,02 | 106,05 | 118,55 | 3,05 | 1,19 | 1,91 | 2,93 | 4,31 | 17,70 | 2,77 | 16,33 | 18,62 | 19,38 |
| León | 16 | 84,99 | 21,14 | 61,87 | 92,17 | 103,14 | 3,08 | 1,21 | 1,92 | 2,91 | 4,35 | 19,93 | 3,27 | 18,45 | 21,61 | 21,93 |
| Lleida | 16 | 140,99 | 52,46 | 85,33 | 144,39 | 188,71 | 3,02 | 1,20 | 1,84 | 2,84 | 4,34 | 29,59 | 4,90 | 26,65 | 31,15 | 33,19 |
| Lugo | 16 | 73,08 | 15,41 | 56,07 | 78,52 | 87,09 | 3,06 | 1,20 | 1,92 | 2,89 | 4,37 | 20,07 | 4,33 | 17,87 | 20,89 | 22,89 |
| Madrid | 16 | 88,05 | 27,11 | 59,71 | 90,11 | 108,47 | 3,07 | 1,19 | 1,93 | 2,91 | 4,35 | 33,79 | 5,86 | 30,85 | 35,65 | 36,87 |
| Málaga | 16 | 113,73 | 50,43 | 60,59 | 114,07 | 160,94 | 3,10 | 1,17 | 2,01 | 2,92 | 4,34 | 19,02 | 2,68 | 18,81 | 20,02 | 20,43 |
| Murcia | 16 | 107,57 | 32,88 | 71,37 | 111,70 | 136,50 | 3,09 | 1,17 | 1,99 | 2,94 | 4,35 | 21,52 | 3,36 | 20,33 | 22,57 | 23,18 |
| Navarra | 16 | 139,86 | 40,06 | 97,19 | 148,17 | 174,95 | 3,05 | 1,19 | 1,89 | 2,88 | 4,29 | 30,88 | 4,81 | 28,90 | 32,48 | 33,60 |
| Orense | 16 | 73,21 | 14,59 | 57,04 | 80,84 | 86,36 | 3,07 | 1,20 | 1,92 | 2,90 | 4,38 | 18,59 | 3,54 | 16,47 | 19,68 | 20,82 |
| Palencia | 16 | 89,41 | 23,05 | 64,70 | 96,01 | 108,03 | 3,07 | 1,20 | 1,92 | 2,90 | 4,35 | 24,18 | 4,14 | 21,85 | 25,57 | 26,43 |
| Pontevedra | 16 | 78,35 | 21,73 | 53,76 | 84,15 | 99,32 | 3,05 | 1,19 | 1,89 | 2,89 | 4,36 | 20,62 | 3,67 | 19,11 | 21,86 | 22,74 |
| La Rioja | 16 | 87,69 | 22,32 | 63,63 | 90,91 | 106,82 | 3,06 | 1,19 | 1,93 | 2,90 | 4,37 | 26,56 | 4,36 | 24,49 | 28,18 | 28,94 |
| Salamanca | 16 | 84,75 | 24,26 | 57,80 | 94,72 | 106,90 | 3,08 | 1,20 | 1,92 | 2,92 | 4,29 | 19,64 | 2,85 | 18,45 | 20,62 | 20,98 |
| Segovia | 16 | 86,75 | 25,42 | 58,20 | 91,45 | 109,43 | 3,08 | 1,20 | 1,93 | 2,91 | 4,36 | 22,86 | 3,11 | 22,15 | 23,92 | 24,44 |
| Sevilla | 16 | 86,20 | 38,24 | 42,25 | 95,66 | 120,93 | 3,05 | 1,19 | 1,90 | 2,90 | 4,28 | 20,76 | 3,22 | 19,61 | 22,07 | 22,61 |
| Soria | 16 | 83,82 | 20,09 | 61,58 | 89,94 | 99,13 | 3,11 | 1,20 | 1,96 | 2,96 | 4,39 | 23,91 | 3,88 | 21,42 | 25,48 | 26,35 |
| Tarragona | 16 | 115,17 | 41,05 | 71,29 | 112,86 | 153,18 | 3,11 | 1,16 | 2,01 | 2,94 | 4,40 | 30,05 | 4,15 | 27,91 | 31,13 | 31,66 |
| Teruel | 16 | 89,73 | 21,70 | 65,42 | 95,51 | 108,95 | 3,09 | 1,21 | 1,95 | 2,94 | 4,40 | 25,17 | 3,99 | 23,31 | 26,90 | 27,96 |
| Toledo | 16 | 97,11 | 30,58 | 62,88 | 103,27 | 124,87 | 3,07 | 1,18 | 1,94 | 2,90 | 4,35 | 19,55 | 2,65 | 19,44 | 20,30 | 20,93 |
| Valencia | 16 | 98,39 | 31,02 | 64,17 | 102,72 | 123,52 | 3,01 | 1,18 | 1,88 | 2,86 | 4,30 | 23,34 | 3,54 | 21,68 | 24,57 | 25,53 |
| Valladolid | 16 | 85,01 | 24,65 | 58,01 | 89,36 | 105,15 | 3,05 | 1,20 | 1,90 | 2,88 | 4,36 | 24,50 | 4,12 | 22,49 | 26,03 | 26,62 |
| Zamora | 16 | 79,91 | 19,41 | 58,84 | 87,73 | 95,88 | 3,06 | 1,20 | 1,91 | 2,89 | 4,33 | 18,27 | 3,48 | 16,13 | 19,31 | 20,69 |
| Zaragoza | 16 | 94,87 | 26,92 | 64,64 | 99,35 | 118,11 | 3,05 | 1,18 | 1,91 | 2,90 | 4,35 | 26,60 | 4,46 | 24,66 | 28,25 | 29,05 |

Note:

*Per capita sales are measured in packs of 20 cigarettes per year. The price is measured in real euros of 2016. GDP per capita is expressed in thousands of real euros of 2016.

### 2.2 Empirical methodology

Data-driven anomaly detection systems have been discussed in the literature as distortions detection systems in many fields of application (see[20-23]). Such systems aim to detect any abnormal deviations from the normal observations of any data set. Therefore, these methodologies provide a good opportunity to detect anomalies in tobacco sales. Furthermore, given the above characteristics, Spain seems a reasonable candidate country to quantify anomalies.

The aim of this work is the tobacco sales anomaly detection at province geographic level. A prediction of the upper and lower limits of tobacco sales at the provincial level is proposed as a methodology in order to identify any abnormal deviation from this behavior in tobacco sales. In this way, the methodology is proposed through a supervised learning method, adjusting a model to tobacco sales as a dependent variable from price and GDP as independent variables. On the other hand, the detection and estimation of anomalies is done through an unsupervised method, as mentioned before by means of the computation of upper and lower intervals. Several statistical and machine learning models were compared for finding the best model for predicting the tobacco sales of each province (these methods/models are presented in this section).

The main methodology consists on splitting the data into a training and test set of all the available province for the Spanish territory, where the training set consist on all the province available with the exception of the province to predict, which is on the test set. In other words, all province tends to be used to predict a chosen province without including the predicted one. As is common to explain the behavior of tobacco consumption in Spain (see [24-26]), the dependent variable is the per capita tobacco sales for every province and the independent variables are the per capita GDP and price:

$$Tobacco\ sales = f(price, GDP, Pop^{18+})$$

To model the relationship between the dependent variable and the independent variables (the characteristic vector x), two supervised learning methods have been used. In addition, in order to estimate the upper and lower limits of the prediction interval, quantile predictions will be used as intervals following the methods explained in this section.

The first method used to model the relationship between variables is Quantile Regression (QR). This method was introduced by[27] for the estimation of models in which the quantiles of the response are modelled to depend on the features. The τth quantile for a population is the sample where the $100/\tau\%$ proportion of the population lies. This model relationships between different quantile predictors and the dependent variable, in this case gives a good interpretability of anomaly detection results as is possible to identify an anomaly within a given range(see[23]).

The conditional α-quantile $q$ of a scalar variable Y, $P(Y \leq q|I) = \alpha$ where the probability *0< a<1* is given and *I* denotes an information set generated by independent variables X. For a complete justification of the method, (see[27]).

For the purposes of this work, two models were combined to build the intervals for detecting anomalies. This is, for the conditional 0.1-quantile as a lower interval and 0.9-quantile as an upper interval, for every province. By construction the probability that a value belongs to the interval between the upper and lower interval is:

$$P( l<=X <= u) = P(X <= u) - P(X <=l)=0.9–0.1=0.8$$

In contrast to the method of least squares that estimates the conditional mean, this method is based primarily on choosing a model for the conditional quantile. Depending on the strength of the assumptions imposed, a range of parametric or non-parametric options are available, (see[28]).

For assessing the models, the conditional median response for each province was modelled, which means the 0.5-quantile. Not only the models are evaluated for punctual predictions but also the intervals for choosing the best model with a good performance in both tasks, this is discussed later in this section.

A bagging method is proposed in this work as an approach for estimating conditional quantiles. A combination of Random Forest (RF) and QR where proposed by[29] giving as a result Quantile Regression Forest (QRF) approach. One of the main differences between RF and QRF is that QRF for each node of each tree maintains the values of all observations of the node, but RF only maintains the mean of the observations found in the node, (see[29]). Ranger is a fast implementation of RF or recursive partitioning, (see[30]), particularly suited for high dimensional data.[3]

To detect anomalies, two methods were selected in order to build PIs trough conditional quantiles, for every new observation of the response variable there is a high probability that it lies within the prediction interval (PI), (see[24]). Furthermore, an anomaly detection and quantification system is proposed by using an upper interval and lower interval computed through the fitted models.

As mentioned before the PIs are computed through the calculation of the chosen conditional 0.1-quantiles and 0.9-quantile for lower interval and upper interval respectively. One requirement for the decision of α for the intervals is using a symmetric range (i.e you can't use the 0.1 quantile as the lower interval and the 0.7 quantile as the upper). It is not interest for this research to find the better intervals within a model but to provide a methodology for computing the ratios of abnormalities as shown on section 5.4.

---

[3] The R libraries *ranger* was used to fit a QRF model respectively with the default settings.(see[31])

The proposed method assumes a uniform distribution with endpoints as the lower and upper limits of the computed PIs. Every point outside this interval range is considered as abnormal, the intervals are also used to quantify the ratio of abnormality for the response variable.

To remark, the observed response for province p is abnormal if either case is true:

$$\begin{cases} y_i > y_t \\ y_i > y_{(100-n)} \end{cases}$$

where $\eta(0 \leq t \leq 100\%)$ presents the chosen quantile level being this symmetrical, limits y (100 −t)% level and y t% level represent the upper and lower conditional quantiles, respectively. A small chosen value of t will lead to a larger number of provinces predicted as abnormal.

For training the model a data partition was performed, as explained in the aforementioned sections, the predictive accuracy of the models was measured by splitting the data into training and test sets.

The error assessment was performed either by using a 0.5-quantile prediction for the quantile versions, the interval prediction were used to determine the quantity of the abnormality of tobacco sales that was evaluated with the results of surveys and with some metrics to assess the quality of this intervals.

The performance of the predicted responses ($\hat{y}_i$) in relation to the observed responses ($y_i$) of the training and test set were assessed by computing the following error metrics:

**Table 2.** Averaged error metrics for the fitted models at the training set.

| Averaged error metric | Calculation way |
|---|---|
| Prediction error | $e_i = y_i - \hat{y}_i$ |
| Mean squared error | $MSE = \frac{1}{n}\sum_{i=1}^{n} e_i^2$ |
| Mean absolute error | $MAE = \frac{1}{n}\sum_{i=1}^{n} |e_i|$ |
| Mean absolute percentage error | $MAPE = \frac{1}{n}\sum_{i=1}^{n} |\frac{e_i}{y_i}|$ |
| Median Absolute Error | $MedAE(y, \hat{y}) = median(|y_i - \hat{y}_i|, ..., |y_n - \hat{y}_n|)$ |
| Median Squared Error | $MedAE(y, \hat{y}) = median((y_1 - \hat{y}_1)^2, ..., (y_n - \hat{y}_n)^2)$ |
| Median Absolute Percentage Error | $MedAPE(y, \hat{y}) = median\ of\ sorted\ |\frac{\hat{y}_i - y_i}{y_i}|$ |

Note:

* It is computed by ordering the absolute percentage error (APE) from the smallest to the largest and using its middle value (or the average of the middle two values if N is an even number) as the median.

* For further understanding of this formulas and its statistical properties, (see[32]).

In addition to evaluating the punctual prediction, the prediction interval made has also been evaluated in this work. The academic literature has placed special emphasis on point prediction with respect to interval predictions and predictive densities, consequently there has been little work on the evaluation of PI, (see[33]).

A review of evaluating point prediction methods used at this work is in the Table 2, where the selected metrics assesses the accuracy into the train and test set. However, as the main idea of forecasting is in decreasing the uncertainty, an interval prediction evaluation is performed as well. Table 3 summarizes the metrics used to evaluate the prediction interval.

**Table 3.** Prediction interval accuracy for the fitted models at the training set.

| Accuracy metric | Calculation way |
|---|---|
| Mean Internal Score | $MIS = (p_u - p_l) + \frac{2}{\alpha}(p_l - y)1(y < p_l) + \frac{2}{\alpha}(y - p_u)1(y > p_l)$ |
| Range | $range = \frac{1}{n}\sum_{i=1}^{n}|\frac{p_{u_i}}{p_{l_i}}|$ |
| Coverage | $covareage = \frac{1}{n}\sum_{i=1}^{n}\left(1(y_i < p_{l_i}) \times 1(y_i < p_{l_i})\right)$ |
| Pinball | $pinball = (1 - \alpha)\sum_{\hat{y}_i < \hat{b}_i}|\hat{y}_i - \hat{b}_i| + \alpha \sum_{\hat{y}_i < \hat{b}_i}|\hat{y}_i - \hat{b}_i|$ |

Note:

*Where pl is the lower PI, pu the upper PI, α is the significance level, y the actual value and $1(\cdot)$ is the indicator function, for more details see[30].

*Where n is the number of sample size, $p_u$ for upper PI, $p_l$ for lower PI and y the actual value for each observation $i$.

*Where $\hat{b}_i$ is the predicted value of a interval (either an upper, or a lower).

*MIS balance *coverage* and *range* of the PI, the best choice is when a model has high coverage, but also short intervals.

*Pinball loss function show how well a quantile capture the data, the lower the value of pinball is, the closer the interval is to the specific quantile of the holdout distribution.(see[34]).

Although anomaly detection has been used in many previous works for other disciplines, the novelty of this method is about complement the abnormality detection with abnormality quantification. As explained on the previously, the computed intervals are used to quantify the ratio of abnormality, the following formulas are applied to accomplish that task:

**Table 4.** Quantification of anomalies in per capita tobacco consumption.

| Anomaly ratio | Calculation way |
|---|---|
| Upper anomaly ratio (UAR) | $UAR = \frac{y_i - p_u}{p_u}$ |
| Lower anomaly ratio (LAR) | $LAR = \frac{y_i - p_l}{p_l}$ |

Note:

*Where pl is the lower PI and pu the upper PI.

As a final product of this work a model could discern between a province with abnormal tobacco sales and a province without abnormal tobacco sales, but also, when abnormality is detected this could abnormality due to a quantity is under a lower interval or abnormality due to a quantity is over a upper interval.

### 3. Results

The results of this article are shown in three parts. First, the evolution over time of the anomalies detected is shown (both the provinces in which sales are lower than fair values and those in which sales are higher than expected). Second, the temporal evolution of the regional anomalies detected in Spain is shown. Finally, the geographical distribution of the anomalies detected is shown.

As indicated, to quantify the anomalies in the Spanish provinces we will use the upper anomaly ratio (UAR) and the lower anomaly ratio (LAR). The average UAR and LAR for the Spanish territory have been represented, averaging the prediction ratio of tobacco sales per capita below the lower limit (the lower prediction interval) and also for the prediction ratio of tobacco sales per capita above the upper bound (the upper prediction interval). Figure 2 shows the aforementioned index and in it two important questions can be observed: (i) the magnitude of the average upper anomaly exceeds 40%, while the average lower anomaly does not reach 15%, (ii) the temporal evolution shows a upward trend in the lower anomaly and descending in the upper anomaly.

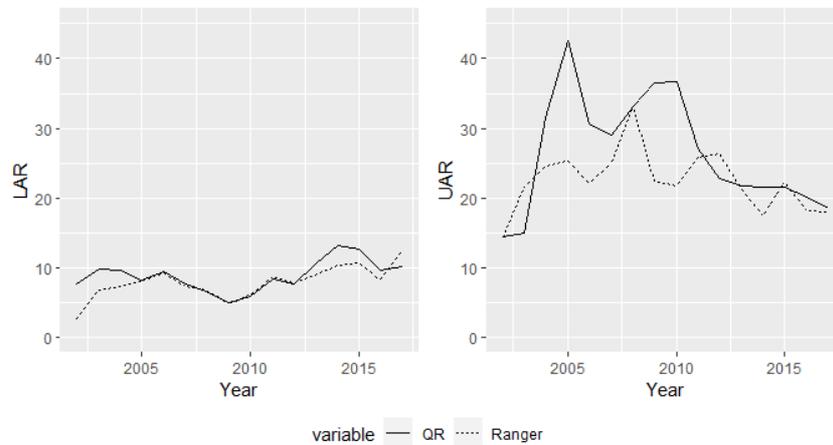

Figure 2. Average UAR and LAR for the Spanish territory.

On the side of the magnitude of LAR and UAR, although, as will be seen in this section, the geographic pattern plays a key role, it seems that there is a significant difference between both indices. As for the average LAR, it represents the average percentage of those provinces that present observed sales below the estimated values. Therefore, given that this index is conditioned by crossborder and illegal trade, it seems that both activities have remained constant during the period studied. However, the average LAR has increased notably until 2005, showing a decreasing trend since then. The LAR index represents the average anomaly of the provinces in which the observed sales exceed those estimated by the model. For this reason, both the effect of tourism and that of the crossborder with countries where tobacco is more expensive than in Spain, determine the magnitude of the LAR index. In this line, according to [16], the border provinces with France have been considered a crossborder effect, while the rest are provinces with a high influence of tourism. Figure 3 shows the part of the average UAR that represents crossborder and tourism. Both effects seem to have a decreasing trend, the crossborder trend being more accentuated. This last highlight is consistent with the recent evidence on crossborder transactions between Spain and France (see [11]).

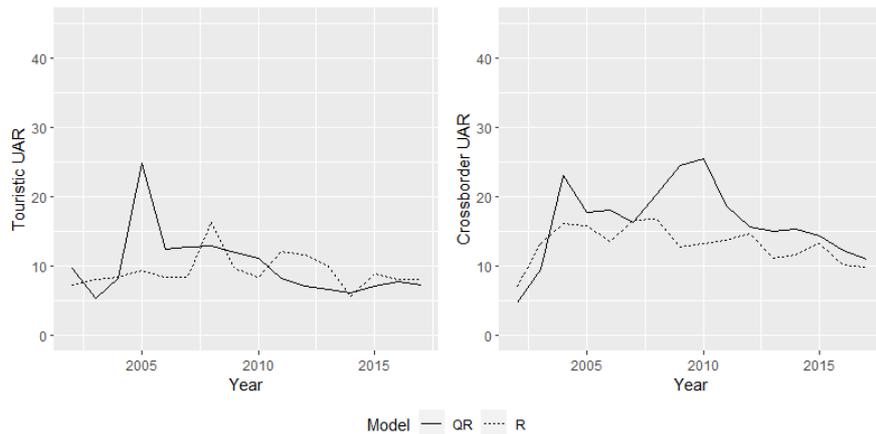

Figure 3. Touristic and crossborder UAR in Spain.

Once the magnitudes have been exposed at a global level, it is interesting to analyze the temporal evolution of UAR and LAR in the Spanish provinces. First, regarding the UAR, there are 6 provinces that stand out for their behavior. As shown in Figure 4, there are three provinces in the south of Spain (Sevilla, Cádiz and Córdoba) in which the UAR is observed for the first time in 2010 and grows notably until 2017, reaching values close to 40% in some cases. On the other hand, in three other provinces (Orense, Pontevedra and Lugo) the UAR has a decreasing trend, in addition, it rarely takes values close to 20%. In addition, Figure 4 also shows the values shown by the EPS performed by the TTCs from 2012 to 2017. As the aforementioned figure shows, in line with the previous literature, it seems that the use of independent data provides estimates of the illicit market lower than EPS. Specifically, while in Cádiz the EPS is slightly oversized, in Córdoba there are substantial differences.

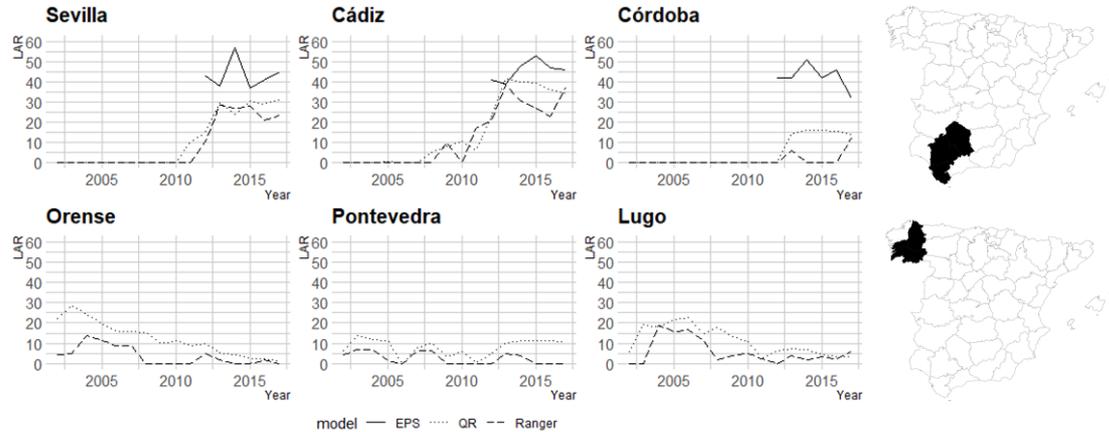

Figure 4. Temporal evolution of UAR in the Spanish provinces.

Six provinces also stand out in the temporal evolution of the UAR in the Spanish territory. On the one hand, Malaga, Alicante and the Islas Baleares, provinces with a high influence of tourism[4], present a similar trend. On the other hand, in the border provinces with France, it is observed that in Gerona, Huesca and Guipúzcoa the UAR shows a decreasing trend, with the anomalies detected in Gerona of a much higher magnitude. As indicated in the introduction, Gerona is the Spanish region in which sales fell the most due to border closures due to the public health crisis of COVID-19. Therefore, the results are consistent with what is indicated.

---

[4] According to data from the National Institute of Statistics, the airports in these three provinces are the ones with the most arrivals, after Madrid and Barcelona, the largest provinces in Spain.

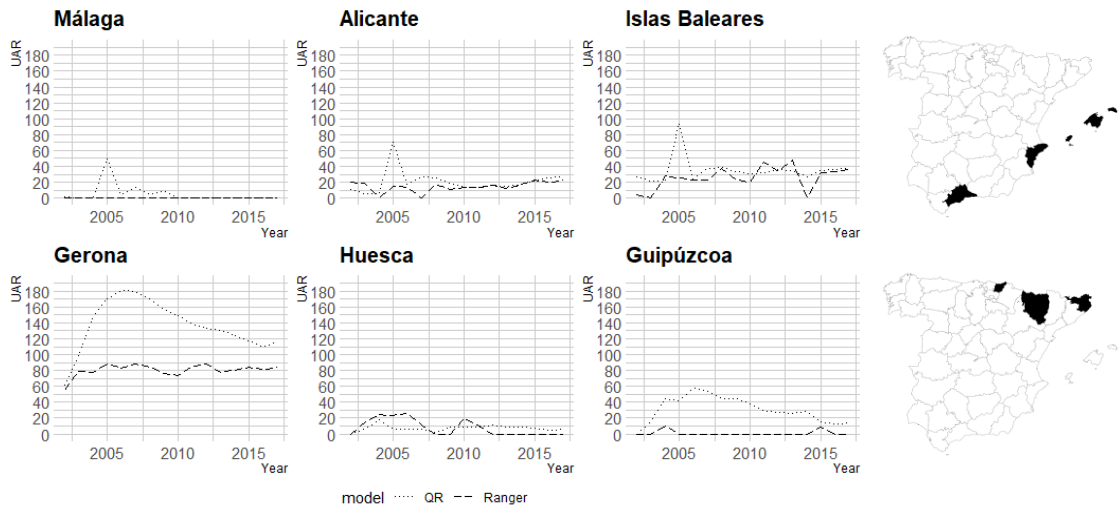

Figure 5. Temporal evolution of LAR in the Spanish provinces.

Although the provincial evolution provides important information, the geographical distribution of the anomalies helps to understand the "contagion effect" of the UAR and the LAR, as well as the behavior of consumption at the borders with other countries. In this sense, Figures 6 and 7 show the geographic distribution of LAR and UAR according to the QR model, respectively. As can be seen, the anomalies in the provinces in which lower-than-estimated sales have been observed reach up to 35% and are concentrated in the Northwest in 2002 and in the South in 2017, something consistent with the previous literature. In addition, regarding the UAR, the anomalies in the provinces with sales above those estimated reach values of 190%. Finally, while in 2002 they were concentrated in tourist provinces and bordering Portugal, in 2017 the tourist provinces remain the same, but it is the border with France where the anomalies are located.

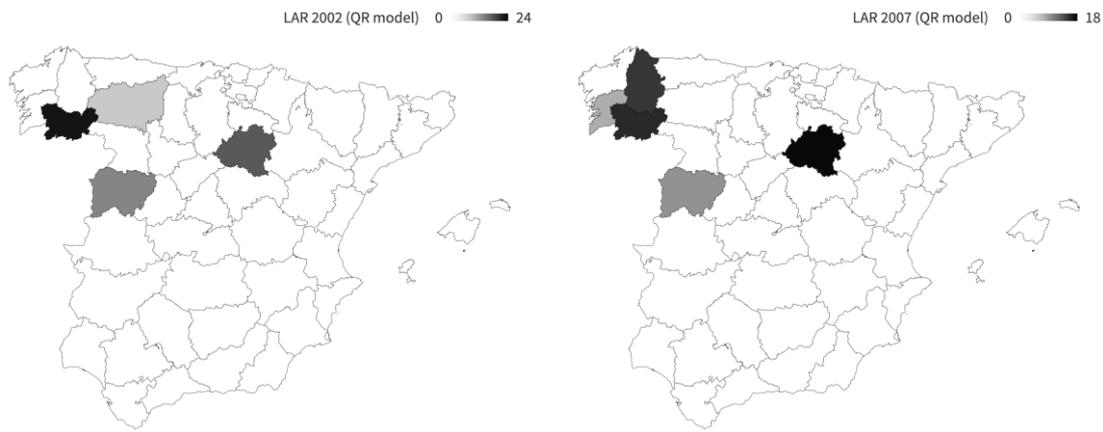

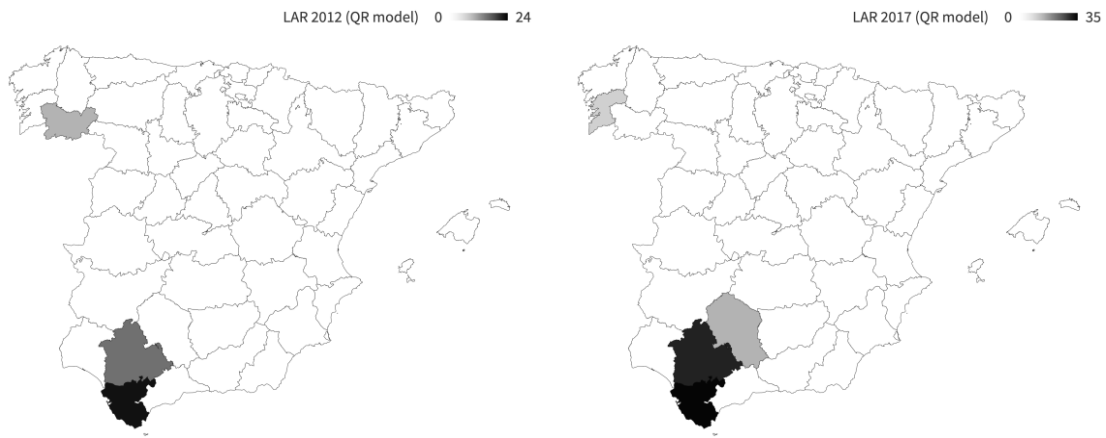

Figure 6. Geographical distribution of LAR in the Spanish provinces (QR model).

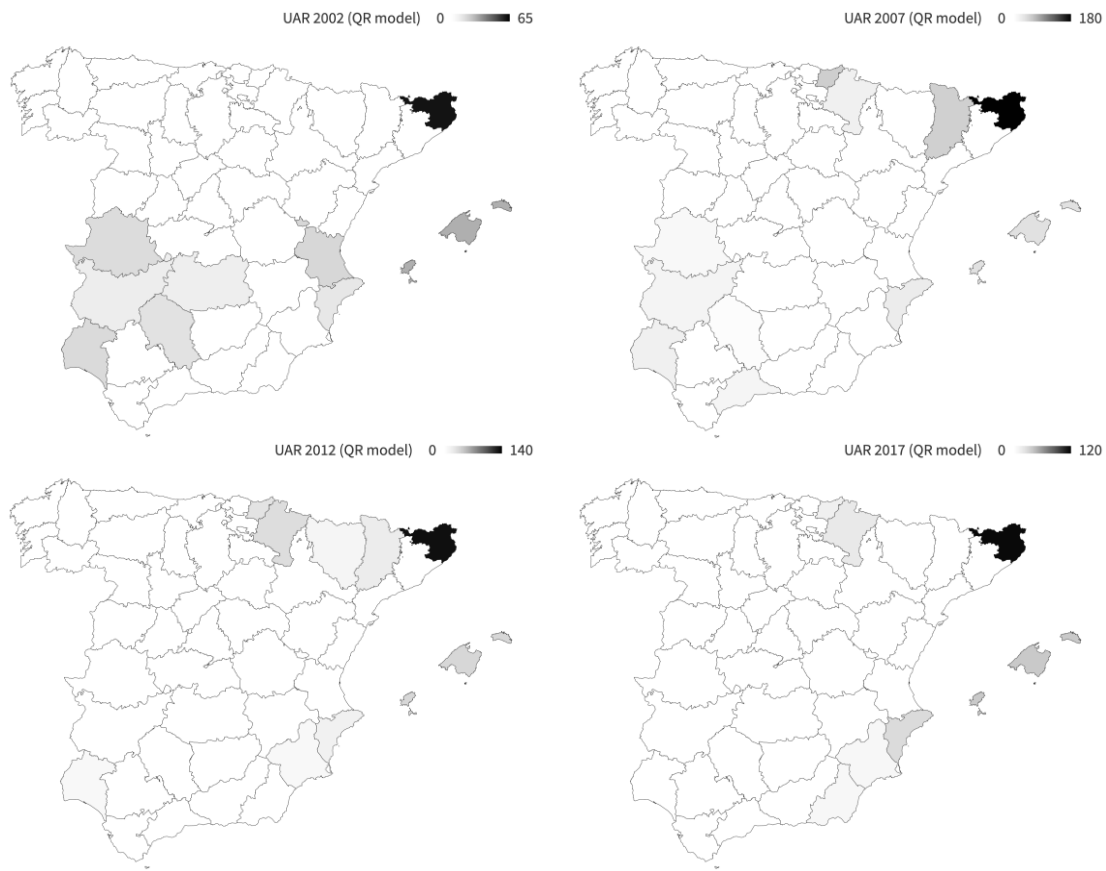

Figure 7. Geographical distribution of UAR in the Spanish provinces (QR model).

One of the main values of this article is the quantification of the LAR and UAR anomalies. Along these lines, the closure of Spain's borders with border countries due to the health crisis caused by COVID-19 during the months of April and May 2020 has made it possible to analyze the robustness of the results of this paper. As shown in Figures 8 and 9, both the geographic pattern and the magnitude of the UAR estimated in this work show robustness with the falls in tobacco sales in Spain in the months of April and May 2020. On the one hand, tobacco sales fell in April and May up to 180 and 160 percent, respectively, against. This magnitude is consistent with the anomalies shown in Figure 7. Furthermore, the geographical pattern is also coincident, with the greatest drops in tobacco sales having been observed in the border areas with France and in the tourist provinces. In addition, the border areas with Gibraltar are those in which tobacco sales have decreased the least during the border closures. It is precisely in these provinces that the highest UARs are observed in the last years analyzed.

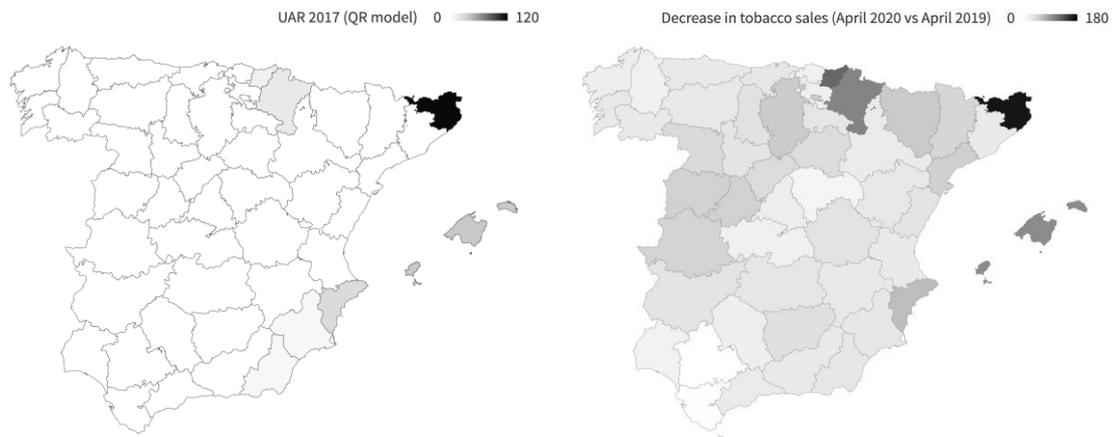

Figure 8. Comparison between the results of the model and the fall in sales of April 2020.

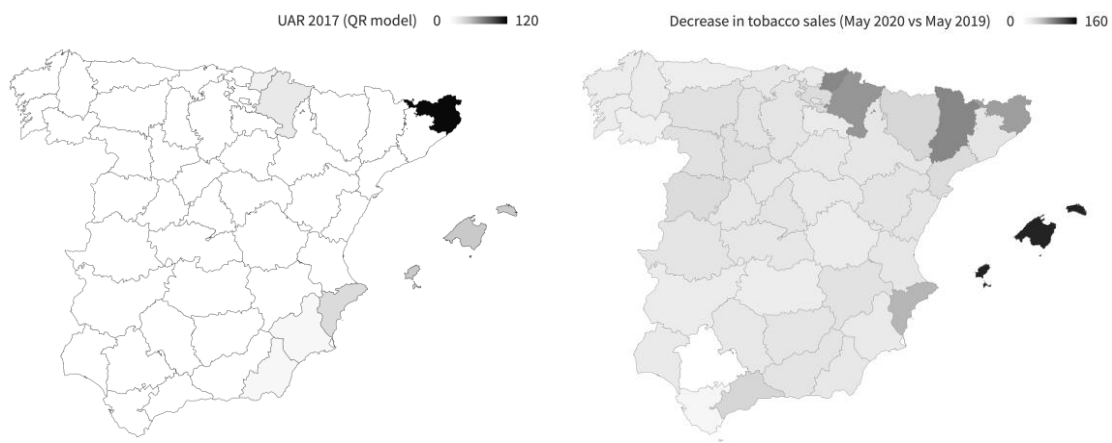

Figure 9. Comparison between the results of the model and the fall in sales of May 2020.

## 4. Conclusions

In recent years there has been a growing interest in knowing the mechanisms that can control cigarette consumption due to the great impact of tobacco consumption on public health. Along these lines, due to the free movement of people and illegal activities, sometimes legal tobacco sales are not a faithful representative of tobacco consumption. For this, there are multiple studies commissioned by the TTCs to demonstrate that there is an illicit and crossborder activity that generates a greater consumption of tobacco in the population than the governments believe. Although there are many initiatives commissioned by TTCs, EPSs are the most widespread studies. These EPSs are in charge of detecting what appears to be illegal trade in provinces where there is less than reasonable tobacco consumption. This study has shown, in line with previous literature, that in Spain the EPSs that are performed to estimate the illicit market (mainly in border areas with Gibraltar) are oversized. In addition, as a contribution to the literature, in this work anomalies have been detected in provinces where sales are higher than fair values, information that TTCs ignore.

To the best of our knowledge, this study is the first to quantify anomalies in regional tobacco sales in Spain, including anomalies in provinces where more than fair values are sold. In particular, the results found on the provinces in which the observed tobacco sales are below fair values are similar to those found in the previous literature: the EPSs overestimate the illicit trade values. In reference to the anomalies detected in the provinces in which tobacco sales above reasonable values are observed, something that is rarely found in the literature, the findings are novel. Specifically, the provinces where tobacco sales are highest relative to fair values are those where sales have fallen the most with border closures driven by the COVID-19 public health crisis. This finding undoubtedly confirms that tobacco sales in Spain are conditioned by the effect of tourism and the price differential with France. Furthermore, it is found that cross-border tobacco purchases between Spain and France show a decreasing trend in recent years, somewhat in line with the previous literature. Although the effect of cross-border tobacco purchases between Spain and France has decreased in recent years, the deviation of sales in the provinces where more tobacco is sold than is reasonable (tourist provinces and the border with France) is much higher on average to deviations from those provinces where sales are below fair values (border with Gibraltar). This result is novel, given that the anomalies of tourist provinces and border provinces with France had never been quantified. Therefore, when the TTCs show results of EPSs in Spain, it must be taken into account that there is an inverse effect to what the EPS detects that must be considered by governments.

Our results show provinces in which smoking control policies cannot be evaluated using official sales, since these sales are altered. In this sense, we find that the provinces in which sales are most affected are border and tourist areas, evidencing the existence of large-scale illegal trade and cross-border purchases. The results support that for some years no anomalies have been found in border areas with Portugal. Therefore, the results reveal the effectiveness of the common policies implemented by the governments of Spain and Portugal, which consists of maintaining a low-price differential between the countries.

All these results provide recipes for the agendas of academics and governments. The academic community should bear in mind that there is more evidence about the overestimation of illicit trade in EPSs and that the average of the excess anomalies is much higher than the average of the default anomalies. In addition, policy makers should consider that there are provinces where the evaluation of the effectiveness of anti-smoking policies cannot be evaluated using official sales. The allocation of resources to control smoking must consider the abnormalities identified in this report. If not, the provinces in which there is excessive consumption distorted will have more recourse to control a tobacco habit that is not real. On the contrary, the border provinces with Gibraltar will have fewer resources to control smoking if official sales are used, when the reality is that there is hidden consumption due to illicit trade.

In conclusion, the results seem to show that EPSs overestimate the value of illicit trade. Furthermore, in Spain, the provinces with sales volumes above fair values have higher ratios than those in which sales are below fair values. Therefore, it seems that the sum of the effect of tourism and cross-border purchases between Spain and France is higher than the cross-border purchases between Spain and Gibraltar detected by the EPS. Finally, the anomalies prevent the Spanish government from knowing the total benefit for public health generated by the policies against smoking.

**Annex. Model error measurements**

**Training set error**

The training set comprises the data without the province to predict for every year, at the following Figure 10, it shows the averaged results (Table 5) of the metrics presented on subsection 2.2.

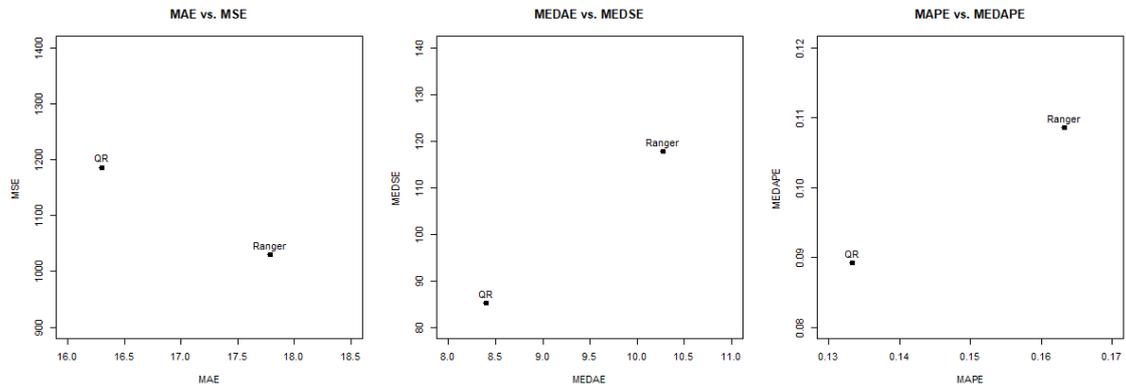
Figure 10. Scatter plots for the error of fitted models at the training set.

As discussed on some research of error measurement (see[29]) different statistical properties reflects every metric, with the errors shown on the previous table QR shows minimum error on training set.

In order to avoid the bias of the averaged metrics, the following density plots are shown in the figure 11. with this we can confirm the superiority of QR over Ranger at the training set.

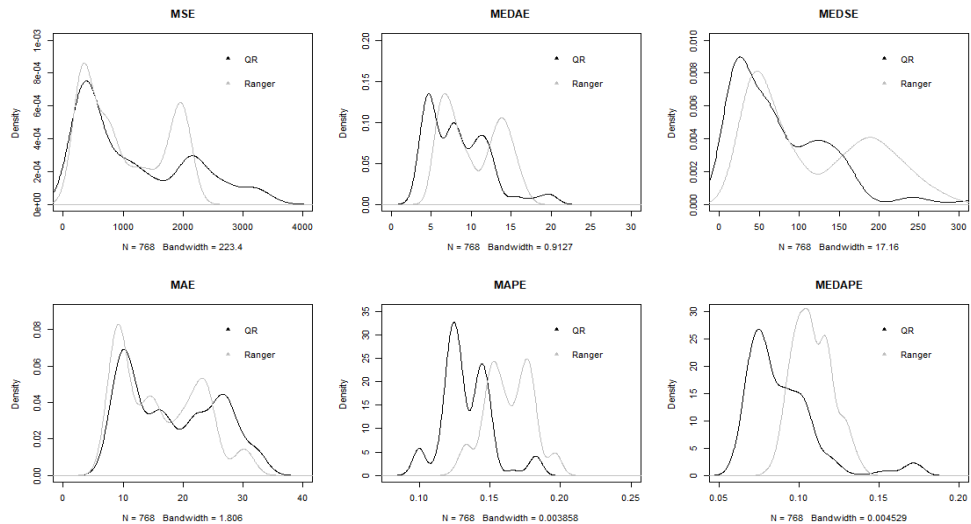
Figure 11. Density plots for the errors of fitted models at the training set.

**Test set error.**

The test set comprises the data with the province to predict for every year, at the Figure 12, it shows the averaged results (Table 6) of the metrics presented on subsection 2.2 are shown.

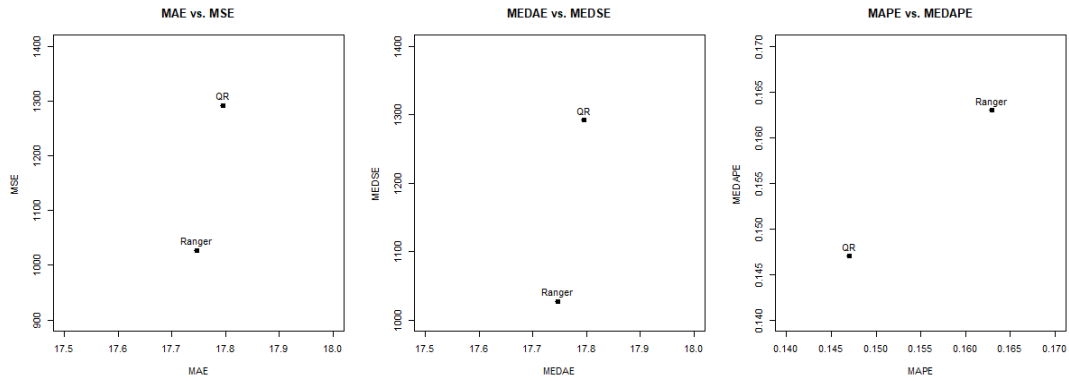
Figure 12. Scatter plots for the fitted models at the training set.

At the test set the predictions errors are very similar, but the square metrics (MSE and MEDSE) penalizes the QR showing a slightly superiority of Ranger.

The density plots shows the distribution of errors at the test set (figure 13), where the similarity of test errors are present with the difference at the MAE Error.

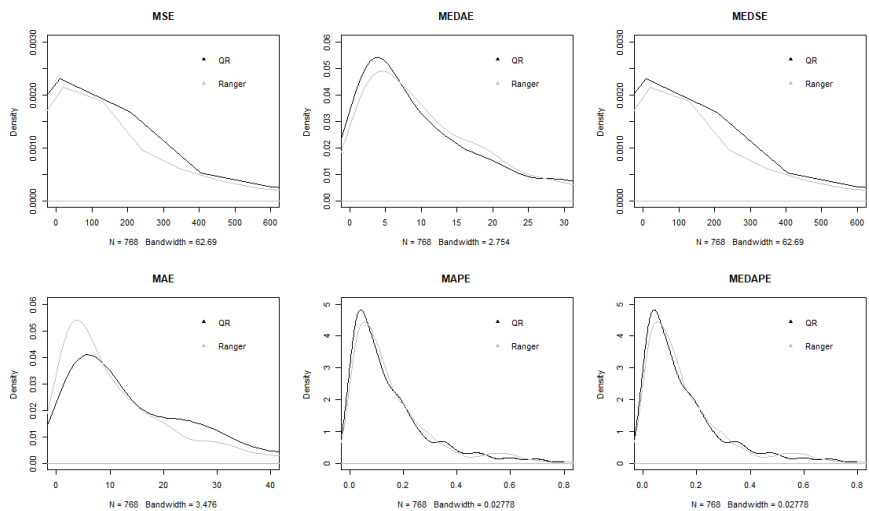
Figure 13. Density plots for errors of the fitted models at the test set.

**Interval Score Metrics.**

This subsection shows at Table 7 metrics for assessing the prediction intervals which are the main novelty use for this work. By using these metrics, a wide overview of how intervals are fitted is potentially used to discard a method for abnormality detection and quantification as this work propose.

The results of Table 7 are shown visually at figure 14, where the superiority of QR over Ranger is present. The MIS and Pinball (average Pinball lw and Pinball hi) shows better performance of the intervals, also the pinball score for bot metrics. The Range and coverage of Ranger are smaller than QR, in this case is the intervals are wide enough to cover the regular points and having a better fit of QR intervals.

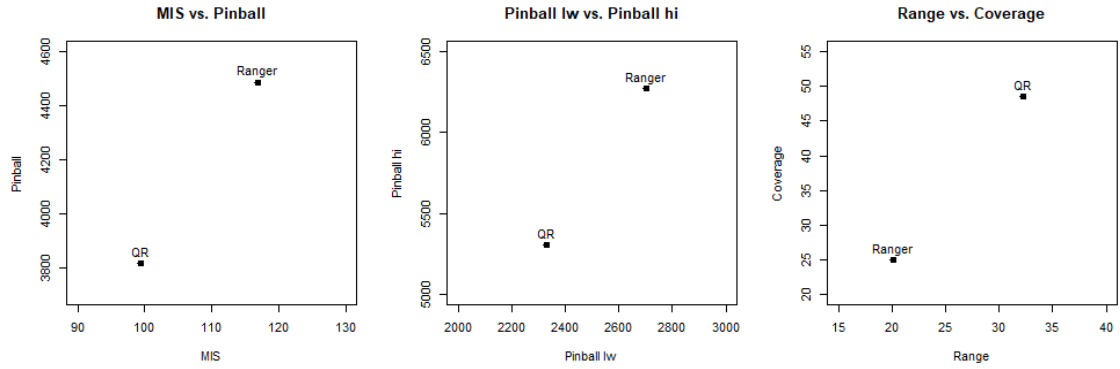

Figure 14. Averaged Interval metrics for the predicted intervals.

| | MAE | MSE | MAPE | MEDAE | MEDSE | MEDAPE | MAE |
|---|---|---|---|---|---|---|---|
| QR | 16,3 | 1185,9 | 0,13 | 8,4 | 85,19 | 0,09 | |
| Ranger | 17,79 | 1029,36 | 0,16 | 10,27 | 117,78 | 0,11 | |

Table 5. Averaged Error metrics for the fitted models at the training set.

| Model | MAE | MSE | MAPE | MEDAE | MEDSE | MEDAPE |
|---|---|---|---|---|---|---|
| QR | 17,8 | 1292,18 | 0,15 | 17,8 | 1292,18 | 0,15 |
| Ranger | 17,75 | 1027,63 | 0,16 | 17,75 | 1027,63 | 0,16 |

Table 6. Average metrics for the prediction at the test set.

| Model | MIS | Coverage | Range | Pinball_lw | Pinball_hi |
|---|---|---|---|---|---|
| QR | 116,81 | 25,04 | 20,14 | 2702,23 | 6269,05 |
| Ranger | 99,41 | 48,52 | 32,22 | 2328,82 | 5305,85 |

Table 7. Averaged Interval metrics for the predicted intervals.